\documentclass{elsarticle}

\usepackage{ecrc}
\usepackage{hyperref}

\usepackage{amssymb}

\usepackage[figuresright]{rotating}

\begin{document}
\begin{frontmatter}




\title{Towards a Digital Twin Modeling Method for Container Terminal Port}


\author[a,b]{Faouzi Hakimi \corref{cor1}} 
\author[a,b]{Tarek Khaled \corref{cor1}}
\author[a]{Mohammed Al-Kharaz}
\author[a]{Arthur Cartel Foahom Gouabou} 
\author[a]{Kenza Amzil}

\address[a]{DMSLOG AI, 5971 W 3rd St, Los Angeles, CA, USA}
\address[b]{Aix-Marseille University, 58 Blvd Charles Livon, 13007 Marseille, France}

\begin{abstract}
This paper introduces a novel strategy aimed at enhancing productivity and minimizing non-productive movements within container terminals, specifically focusing on container yards. It advocates for the implementation of a digital twin-based methodology to streamline the operations of stacking cranes (SCs) responsible for container handling. The proposed approach entails the creation of a virtual container yard that mirrors the physical yard within a digital twin system, facilitating real-time observation and validation. In addition, this article demonstrates the effectiveness of using a digital twin to reduce unproductive movements and improve productivity through simulation. It defines various operational strategies and takes into account different yard contexts, providing a comprehensive understanding of optimisation possibilities. By exploiting the capabilities of the digital twin, managers and operators are provided with crucial information on operational dynamics, enabling them to identify areas for improvement. This visualisation helps decision-makers to make informed choices about their stacking strategies, thereby improving the efficiency of overall container terminal operations. Overall, this paper present a digital twin solution in container terminal operations, offering a powerful tool for optimising productivity and minimising inefficiencies.
\end{abstract}


\end{frontmatter}


\section{Introduction}

As freight volumes continue to surge and operational complexities increase within seaports and container terminals worldwide, the need for efficient and reliable operations has become more crucial than ever before. Amidst this evolving landscape, digital twins have emerged as vital drivers of digital transformation in seaport operations, drawing inspiration from broader industry trends associated with Industry 4.0 \cite{Industry40-maritime}. Encompassing disruptive technologies like artificial intelligence, machine learning, Internet of Things (IoT), cloud computing, and big data analytics, Industry 4.0 promotes digitization, automation, and inter-connectivity \cite{Industry40-1, Industry40-2}.
\\

In the maritime sector, digital twins reflect this trend, holding immense potential to significantly enhance transparency, control, and data-driven decision-making processes \cite{madusanka2023digital}. These virtual representations of physical assets and processes offer unprecedented insights and capabilities to stakeholders. Through the integration of real-time data and advanced analytics, digital twins enable port authorities, terminal operators, and logistics providers to optimise resource allocation, refine workflows, and minimize operational risks. They also serve as valuable resources for scenario planning and predictive maintenance, promoting proactive decision-making and bolstering operational resilience.
\\

Despite these promising prospects, there remains a pressing need to understand the practical application and impact of digital twins within the maritime domain better. This paper presents a digital twin-based approach aimed at optimising the operation of stacking cranes (SCs) responsible for handling containers. Our proposed solution entails creating a virtualised container yard within the digital twin system, which mirrors the physical yard. Utilizing this setup, our approach demonstrates how a digital twin can facilitate the reduction of non-productive moves under two distinct operation strategies. By allowing managers and operators to visualise the operational environment and pinpoint instances of non-productive moves, digital twins contribute to informed decision-making, ultimately enhancing stacking crane optimisation and management.
\\

Subsequent sections elucidate the practical implementation and consequences of digital twins in optimising stacking crane performance in container terminals. In Section 2, a preliminary investigation is conducted through a comprehensive literature review, which contextualises the research within the field of digital twins and maritime activities. The focus is on the operational aspects of terminal yards. In Section 3, we detail the architecture and purpose of the proposed digital twin. A detailed presentation of the tool is given in Section 4. Finally, the proposed solution and the anticipated future work are discussed in Section 5.

\section{Literature review} 
\subsection{Digital Twin}

The digital twin concept, reminiscent of NASA's Apollo project from the late 1960s, involved the ground vehicle closely mirroring its counterpart in space. This approach served both training purposes and the simulation of solutions for critical scenarios \cite{rosen2015importance}. In the early 2000s, the paper \cite{grieves2017digital} introduced the concept of digital twin for product lifecycle management, comprising three primary components: the physical product existing in real space, its virtual counterpart in the digital realm, and the interconnected flow of data and information bridging the two spaces. In 2014, Grieves and Vickers \cite{grieves2014digital} described digital twin as a comprehensive digital representation of a product, covering its structure from the smallest atomic details to its overall shape. Ideally, all information gleaned from inspecting the physical product can be found within its digital twin. This concept has since been broadened from individual products to entire systems \cite{tao2018digital, ding2019defining}. In simple terms, digital twin is a detailed virtual copy of a physical system, where information flows instantly between the virtual and physical versions. It's all about using real-time data analysis, prediction, optimisation, and simulation to create this interconnected system.
\\

Digital twins have gained significance in both the port industry and research, as noted in a study by \cite{PortTechnology2021} and research conducted by Madusanka et al. \cite{madusanka2023digital}. Digital twins are increasingly seen as vital components of future ports and the maritime sector, as they enhance efficiency and streamline logistical processes. 
By providing a unified platform for analysing data from multiple sources, they help researchers and practitioners gain deeper insights into terminal port behavior, uncovering hidden patterns. 
They allow for simulating and analysing system behavior under different conditions, aiding in the exploration of new designs and solutions while preventing potential terminal real problems. 

\begin{figure}[!htbp]
\vspace*{4pt}
\centerline{\includegraphics[scale=0.3]{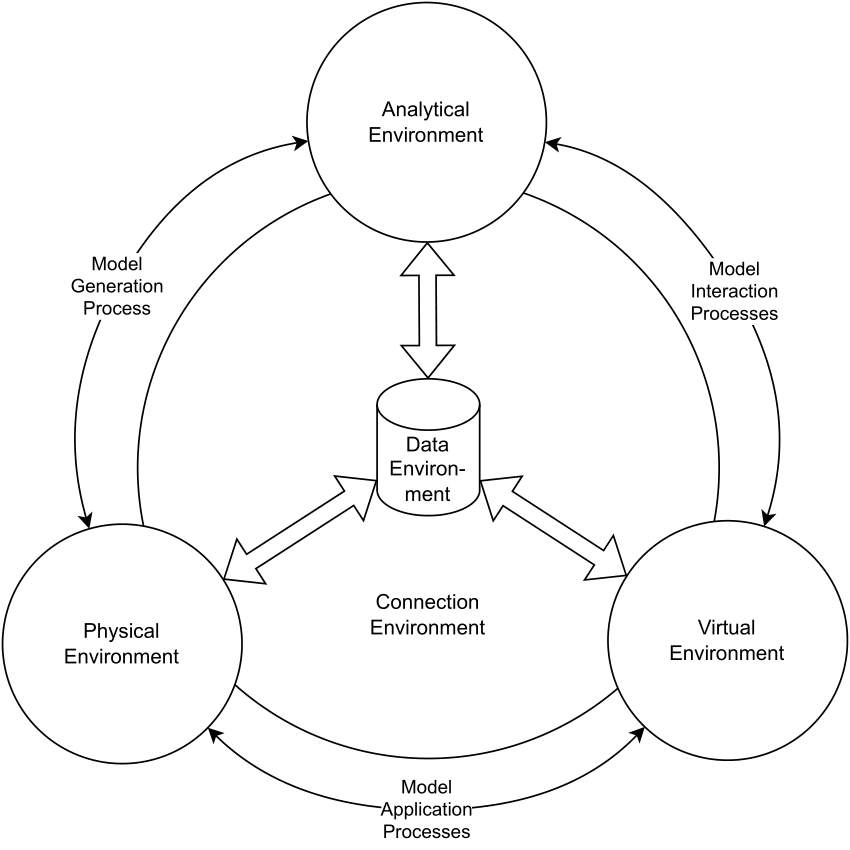}}
\caption{Diagram representing the key components of a digital twin. \cite{DT_diagramm}}
\label{fig:Generic_DT}
\end{figure}

Figure \ref{fig:Generic_DT}  depicts a generic representation of a digital twin, illustrating the flow between four key components: the analytical, physical, virtual, and connection environments, which are mediated by the data environment. It emphasises the continuous feedback loop between real-world and virtual systems for dynamic analysis and decision-making.

\subsection{Terminal Yard}

\begin{figure}[!htbp]
\vspace*{4pt}
\centerline{\includegraphics[scale=0.6]{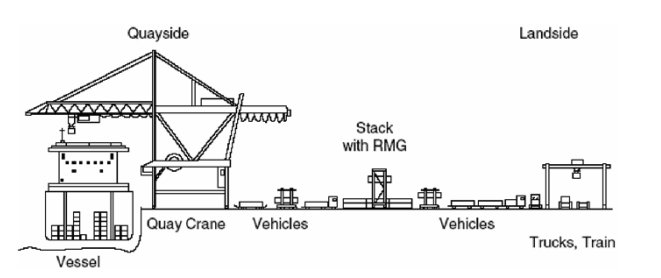}}
\caption{The schematic side view of a container terminal illustrates how the terminal's components are laid out and organised \cite{steenken2004container}. }
\label{fig:side_view_terminal}
\end{figure}

Terminal Yard generally have the same behavior (see Figure \ref{fig:side_view_terminal}), but vary considerably in terms of size and layout. The berthing area is equipped with quay cranes for the loading and unloading of vessels. The containers are stacked in the yard which is divided into a number of blocks. Distinct stack areas are designated for hazardous goods, empty containers and reefer containers, which require electrical power. The truck and train operation area connects the terminal with external transportation networks. 

\begin{figure}[t]\vspace*{4pt}
\centerline{\includegraphics[scale=0.5]{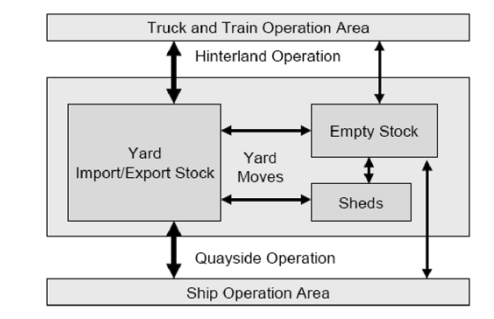}}
\caption{Operation and Transport Flow in a Container Terminal \cite{steenken2004container}}
\label{fig:operation_terminal}
\end{figure}

Figure \ref{fig:operation_terminal} depicts the operational flow within a terminal for both import and export containers. In the case of export containers, upon arrival at the terminal via truck or train, the container undergoes identification and registration processes before being picked up by an internal transport vehicle (ITV) and assigned to a designated storage block within the yard. Stacking cranes then arrange the container within the block according to specified row, bay, and tier parameters. Upon the arrival of the vessel, the container is retrieved from the yard block and loaded onto the vessel at a predetermined stacking position by the quay cranes. Conversely, for import containers, the process begins with the quay crane retrieving the container from the arriving vessel and assigning it to a specific storage block within the yard. Stacking cranes subsequently stack the container within the block, adhering to row, bay, and tier specifications. Finally, upon the arrival of the truck or train, the stacking crane retrieves the container from the yard block and allocates it to the corresponding truck or train for onward external transportation.
\\

A container terminal functions as a dynamic ecosystem, where a multitude of interactions take place between handling, transport and storage units. The inherent complexity of these interactions is compounded by the uncertainty surrounding future events, which amplifies the challenges of terminal management. As a result, real-time planning of logistics operations becomes essential. Activated by specific events or conditions, real-time planning requires rapid decision-making, often within fractions of a second. These decisions include tasks such as assigning transport orders to vehicles, routing and scheduling vehicle routes for landside and dockside transport, allocating storage locations to individual containers, and determining precise schedules and sequences of operations for dockside and stacking cranes.

\section{Architecture and Purpose of the Proposed Digital Twin}
\subsection{Aim of the digital twin}

The use of a digital twin enables the real-time monitoring of each container's journey within the yard, encompassing activities such as stacking, shifting and departure. This provides a more comprehensive understanding, more accurate metric estimation and the capacity to test a multitude of alternative stacking and movement scenarios. Consequently, it is a valuable tool for the development and improvement of a terminal's stacking strategy. 

The primary objective of a stacking strategy is to optimise the expected number of rehandles. To define this quantity more formally, let $c_{kj}$ represent the $j$th configuration of a bay of size $k$. A configuration is here defined as a method of placing the $k$ containers in the bay, assuming that they are all equivalent. Then, according to \cite{KAPHWANKIM1997701} and under reasonable hypotheses (uniform probability of picking for each container, no transfer between bays, etc.), the expected number of shifts following a pick in a bay containing $k$ containers $v_k$ can be expressed as follow:

\begin{equation}
    v_k = \sum_i s_{k_i} \sum_j p_k(i,j)v_k(i,j),
    \label{eq:rehandles}
\end{equation}

with:

\begin{itemize}
    \item $ s_{k_i}$ being the probability that a bay with $k$ containers has the $j$th configuration,
    \item $p_k(i,j)$ being the probability that a bay configuration changes from $c_{ki}$ to $c_{(k-1)j}$,
    \item  $v_k(i,j)$ being the number of rehandles required for the transition from $c_{ki}$ to $c_{(k-1)j}$.

\end{itemize}

Equation \ref{eq:rehandles} illustrates the potential of a digital twin in the planning of a stacking/shifting strategy. The implementation of a digital twin indeed enables the accurate monitoring of the quantities involved in the evaluation of the expected number of rehandles in each bay in the yard $v_k$. 

In the context of a stacking strategy, other container characteristics of interest can be tracked by a digital twin. For instance, the number of days spent on the yard can be an important indicator of its release date. Another metric is the number of rehandles per container, and whether or not its departure has been booked. This information can be essential if the container needs to be shifted before leaving the yard. Information about the port of origin, the customer or the contents of the container can also be tracked, if available.

A digital twin can also be used to monitor other entities, such as vehicles on the yard for instance. Indeed, it may be crucial to consider the distance travelled by the stacking cranes in order to evaluate a stacking strategy. In this regard, a digital twin can be employed to track these machine's position step by step. This monitoring allows to consider this factor when optimising the allocation and reallocation of different elements.

\subsection{Architecture of the Digital Twin}

\begin{figure}[h!]\vspace*{4pt}
\centerline{\includegraphics[scale=0.5]{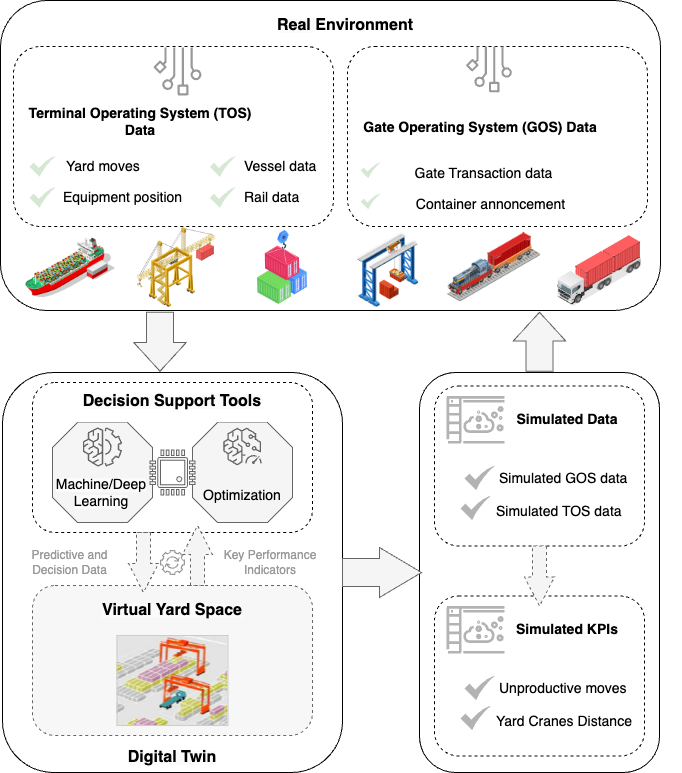}}
\caption{Diagram of the proposed terminal's digital twin.}
\label{fig:Digital_twin_diagramm}
\end{figure}

Figure \ref{fig:Digital_twin_diagramm} provides a schematic representation of the proposed digital twin. This diagram outlines the workflow of a digital twin system for port container terminals, illustrating the data flows and processing steps involved. The system integrates real-time and historical data from Terminal Operating Systems (TOS) and Gate Operating Systems (GOS), which include yard moves, equipment position, vessel and rail data, as well as truck data and container announcements. The obtained data is then processed to generate Key Performance Indicators (KPIs) and predictive data that feed into the digital twin—a virtual model of the port operations.
\\

The digital twin uses this processed data to simulate port activities, generating simulated GOS and TOS data. These data are used to calculate Simulated KPIs, such as unproductive moves and additional simulated TOS data. The objective is to identify inefficiencies and optimise the terminal's operations by simulating different scenarios and predicting outcomes based on the KPIs derived from the data processed.
\\

This virtual replica of the yard therefore serves as the central piece for optimising yard operations in our study. The proposed tool entails the comprehensive capture of yard data, encompassing container positions, attributes such as arrival dates and origins, equipment information, and movement histories. Leveraging this dataset, the digital twin enables the generation of simulated data based on predefined strategies, allowing for real-time comparisons of strategy efficiency against current yard operations. These comparisons are based on KPIs such as rehandling frequency and crane travel distance. Consequently, the digital twin emerges as a dynamic and proactive solution for enhancing yard operational efficiency.

\section{Results}

Data utilised for the proposed digital twin was directly sourced from the terminal operator which we collaborated. To process and simulate this data effectively, we employed Python-based code, which enabled the handling of various yard operations and adherence to terminal rules, including stacking and shift strategies. The interface itself was developed as a web application using JavaScript and PHP, designed to emulate a \textit{TOS-like} view for user convenience. Communication between the backend and frontend components was facilitated through application programming interfaces (APIs), which ensures seamless interaction and data exchange.. \\

The user initially configures the strategy to be tested by the digital twin via a form. The resulting simulated movements are then displayed via the intuitive web interface. These movements are taken directly from the live data of containers entering and leaving the yard. The digital twin interface has been designed to provide users with comprehensive insights and visualisation capabilities. \\

\begin{figure}[t]\vspace*{4pt}
\centerline{\includegraphics[scale=0.4]{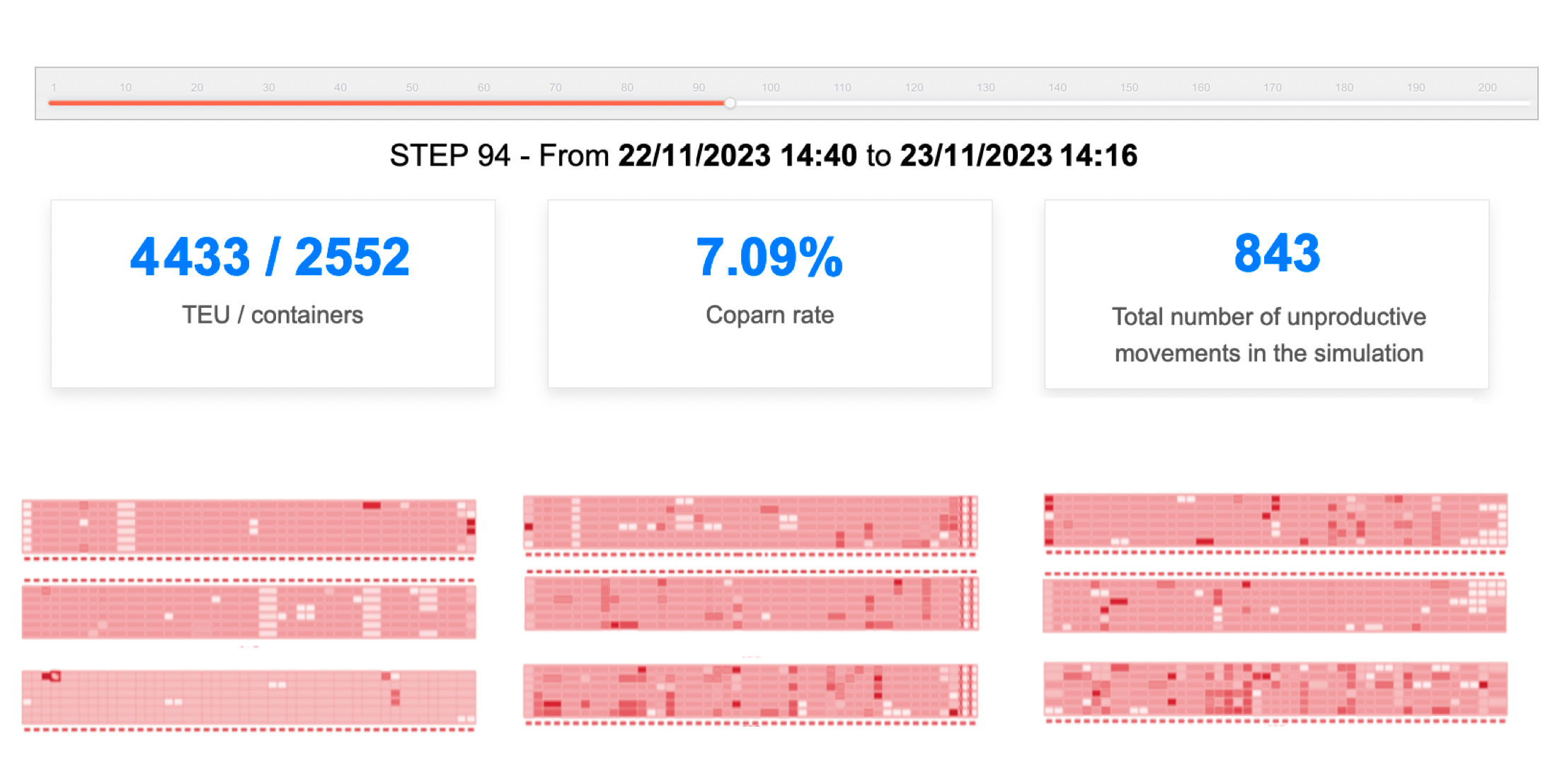}}
\caption{Overview of the terminal and main KPIs. }
\label{fig:terminal_KPI}
\end{figure}

\begin{figure}[t]\vspace*{4pt}
\centerline{\includegraphics[scale=0.35]{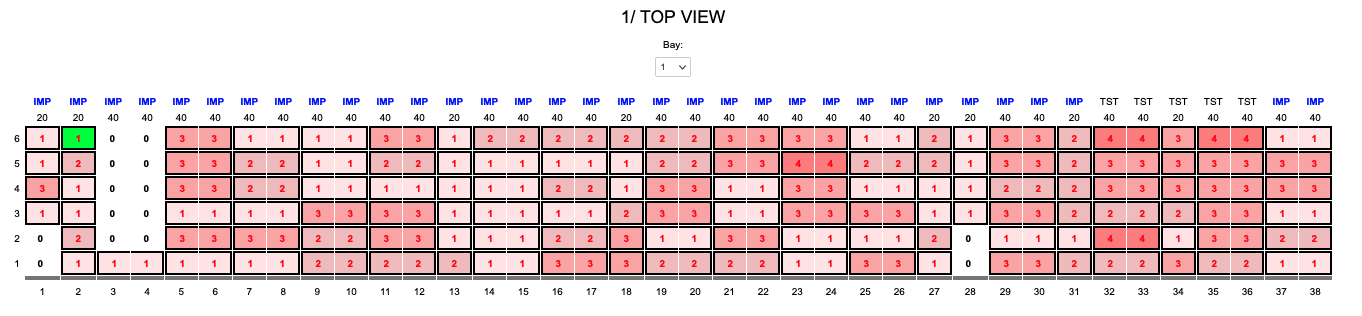}}
\caption{Top view a terminal's block. }
\label{fig:Top_view}
\end{figure}

\begin{figure}[t]\vspace*{4pt}
\centerline{\includegraphics[scale=0.25]{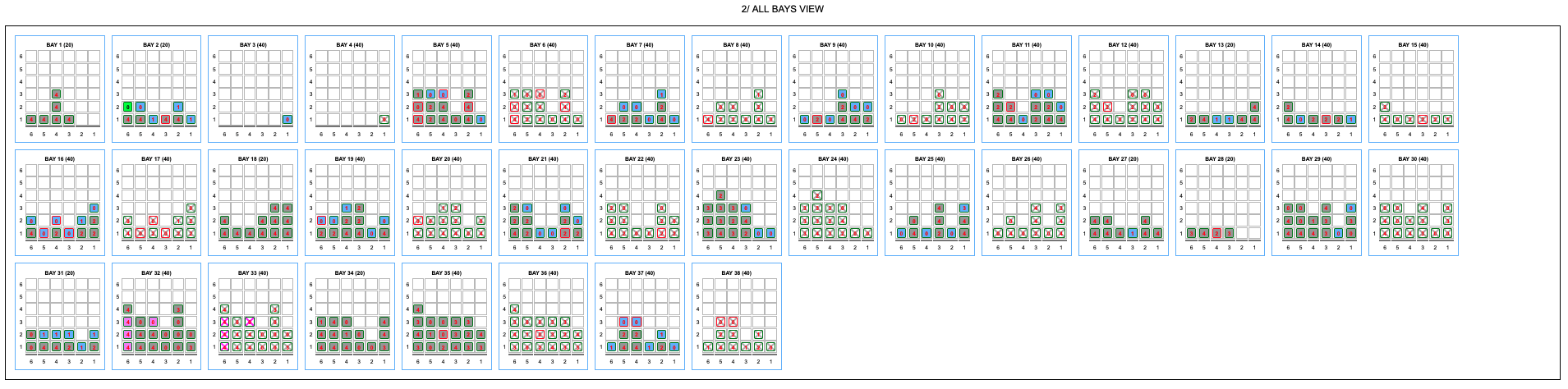}}
\caption{Detailed view a terminal's block. }
\label{fig:global_block_view}
\end{figure}

\begin{figure}[t]\vspace*{4pt}
\centerline{\includegraphics[scale=0.5]{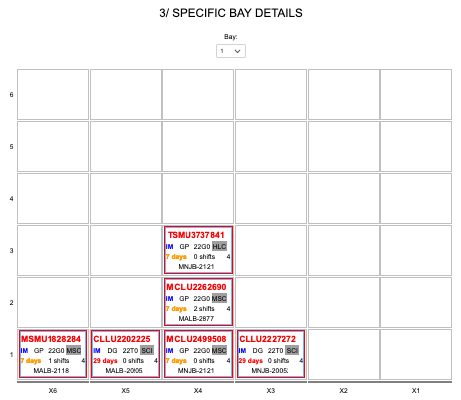}}
\caption{Detailed view of a specific bay. }
\label{fig:stack_detailled_modified}
\end{figure}

At the top of the interface, users can input simulation dates, select the simulation step, and view Key Performance Indicators (KPIs). This layout, depicted in Figure \ref{fig:terminal_KPI}, offers a high-level overview of yard operations. Below this, a top-down satellite view of the yard is presented, illustrating the height of each row. Users can obtain information about the top container in a specific location by hovering over it, enhancing situational awareness and decision-making. This interactive feature is exemplified in Figure \ref{fig:Top_view}. Furthermore, users can access detailed views of each bay within a block (Figure \ref{fig:global_block_view}) and individual bays, providing insights into container characteristics such as days spent in the yard, destination or origin port, and container type (Figure \ref{fig:stack_detailled_modified}). Such detailed visualisations empower users to make informed decisions regarding yard management and optimisation strategies. Notice that the data shown in these figures have been anonymised for confidentiality reasons.

\section{Discussion and Conclusion}

The objective of our study was to introduce a novel digital twin-based approach to optimise stacking crane (SC) operations within container terminal environments. To this end, we developed a virtual representation of the container yard integrated with real-time data feeds, constructing a synchronous system for monitoring, analysing, and controlling SC movements.

The digital twin demonstrated its capacity to accurately replicate the behaviour of physical entities, thereby substantiating its suitability as a testing platform for a range of stacking scenarios. The high degree of congruence observed between the digital twin's output and the corresponding physical counterpart validated the precision and accuracy of the model.
\\

The primary advantage of our digital twin framework lies in its adaptability. Easily configurable settings, modular architecture, and extensible features render the platform suitable for accommodating varied container terminal layouts, equipment types, and operating procedures. Consequently,  the same underlying structure can be customised to suit numerous facilities, ensuring broad applicability without necessitating ground-up construction for every instance.

However, it is important to acknowledge certain limitations to this study. Firstly, the focus of the study was primarily on containers stack, and therefore extending the evaluation to incorporate other operations would yield richer insights regarding the overall interest of the use of a digital twin. Secondly, evaluating long-term effects on safety, productivity, and cost savings requires longitudinal observations beyond the scope of this study.
\\

Future research avenues include expanding the functionality of the digital twin to cover auxiliary tasks, verifying its robustness in large-scale environments over extended durations, quantifying the financial gains derived from reduced downtime and fuel consumption, and exploring innovative means of disseminating pertinent information to end-users. Ultimately, continued exploration of digital twin technology holds great promise for advancing container terminal operations, leading to safer, smarter, and more sustainable practices in the maritime domain.

\bibliographystyle{elsarticle-harv} 
\bibliography{cas-refs.bib}


\clearpage

\end{document}